\documentstyle{amsppt}
\def\pt{\partial}
\def\m{\vert}
\def\ve{\varepsilon}
\def\a{\alpha}
\def\b{\beta}
\def\d{\delta}
\def\g{\gamma}
\def\k{\varkappa}
\def\O{\Omega}
\def\o{\omega}
\def\s{\sigma}
\def\t{\tau}
\vsize=6.6in
\topmatter
\rightheadtext{Asymptotics of perturbed soliton for DS-II equation}
\title{Asymptotics of perturbed soliton for Davey--Stewartson II equation}
\endtitle
\author{R.~R.~ Gadyl'shin, ï.~M.~ Kiselev}\endauthor
\affil{Institute of Mathematics, Ufa Science Centre,
Russian Acad. of Sciences,
112, Chernyshevskii str., Ufa, 450000, Russia}\endaffil
\thanks{This work was supported by RFBR No97-01-00459}\endthanks
\par
\abstract\nofrills{\par
It is shown that, under a small perturbation of lump (soliton) for
Davey--Stewartson (DS-II) equation, the scattering data gain the
nonsoliton structure. As a result, the solution has the form
of Fourier type integral. Asymptotic analysis shows that, in spite of
dispertion, the principal term of the asymptotic expansion for the solution
has the solitary wave form up to large time.}
\endabstract
\endtopmatter
\par
{\bf 0. Introduction.} The Davey--Stewartson II (DS-II) equation,
describing the interaction
of the gravitational and capillary waves on a surface of liquid
\cite{1,2}, is the example of the 2+1-dimensional equation integrable
by the  inverse scattering transform (IST) \cite{3-7}. One of the
most important advantage of IST is that the problem of the structure for
the solution of the original equation is reduced to study spectral data
for associated scattering problem.
\par
The presence of solitary waves in the solution of the DS-II equation is
determined by the existence of poles of the solution for the scattering
problem \cite{6,7}. Since, the problems of soliton stability and
of variation of their parameters are reduced to study dependence
of spectral data with respect to perturbation. For 1+1-dimensional
integrable equations, it is known \cite{8-16}, that  a small perturbation
of the potential for the scattering problem implies a
reqular variation of spectral data and, hence, small changes of soliton
parameters.
 \par
Investigation of solutions for 2+1-dimensional nonlinear equations shows
that there is a more rich collection of forms of evolution. In first,
the 1-dimensional solitons of the Kadomtsev--Petviashvili equation
and of the 2-dimensional nonlinear Schr\"odinger equation are unstable
with respect to transversal per\-tur\-ba\-ti\-ons \cite{4}, \cite{5},
\cite{10}, \cite{17--20}.
One more interesting result is the self-focusing (blow-up), i.e. there are
solutions destroyed at finite time  \cite{21-24}.
\par
In this work the problem of perturbation for the 2-dimensional soliton of
the DS-II equation is considered. It would be naturally to expect results
similar to results for perturbation of solitons for the 1-dimensional
integrable equations. However, the structure of the perturbed solution
proves to be essentially different. For construction of the asymptotic
solution, we use the  IST formalism.  It is shown, that under
perurbation of the 1-soliton initial data, the pole of the solution, for
scattering problem, disappears. That is, the scattering data
gain the nonsoliton structure and the soliton is
unstable with respect to perturbation  of the initial data,
with IST point of view\cite{25}.
On the other hand, the such disappearance of the pole essentially affects
to the asymptotics of the continious component of the scattering data, which
was absent in the unperturbed case. In turn, the constructed
asymptotic expansion of the scattering data is used for the construction
of the asymptotics for the solution of the DS-II equation.
It is remarkable, that, in spite of the nonsoliton structure,
the solution conserves  the soliton form for
the principal term of the asymptotics
up to large time.
\medskip
\par
{\bf 1. Setting of a problem.}
The DS-II equation system is considered in the form \cite{7}:
$$
\eqalign{
&
i\pt_t q+2(\pt^2_z+\pt^2_{\bar z})q+(g+\overline g)q=0,\cr
&
\pt_{\bar z}g=\pt_z\m q\m^2,}
\tag 1.1
$$
where $z=x+iy$, and the overbar represents complex conjugation.
\par
The scattering problem for  (1.1) has the following form \cite{3-7}:
$$
\align
&
\left(\matrix \pt_{\bar z}& 0\cr
                                0 & \pt_z
                        \endmatrix\right)\phi
=\left(\matrix 0 & {q\over2}\cr
-{\overline{q}\over2} & 0\endmatrix\right)\phi,\tag1.2\\
&
E(-kz)\phi(k,z,t)\to \bigg(\matrix 1\cr
                                0\endmatrix\bigg),
\qquad |z|\to\infty,\tag1.3
\endalign
$$
where $E(kz)=\hbox{\rm
diag}(\exp(kz),\exp(\overset{\hrulefill}\to{kz}))$.
Hereafter, the dependence of functions on  $\overline z$ is omitted.
\par
The scattering data  ${\Cal L}(t)$ of the problem (1.2), (1.3) consist of
the continious and discrete parts. The first of them is defined by
the equality
$$
b(k,t)={i\over 4\pi}\int\int_{\Bbb C} dz\wedge d\overline{z}\,
\overline{q\,\phi^{(1)}}
\exp(-\overset{\hrulefill}\to{kz}),
\tag 1.4
$$
where $dz\wedge d\overline{z}=2i\,dx\,dy$ and
$\phi^{(1)}$ is the corresponding component of the vector  $\phi$.
The discrete part  ${\Cal L}^d$  of the scattering data
consists of
poles of $\phi$ and of some their characterizing constants. This part
corresponds to the existence of the solitons and their parameters.
\par
The case, when the continious component vanishes, corresponds to
the pure  soliton solution. The elementary example is the 1-soliton
solution (\cite{7}):
$$
\eqalign{
q(z,t)
&=
q_0(z,t)={2\overline\nu\over|z+4ik_0t+
\mu|^2+|\nu|^2}
\exp\{k_0z-\overline{k_0z}+
2i(k_0^2+\overline k_0^2)t\},
\cr
g(z,t)
&={-4\overline{(z+4ik_0 t+\mu)}^2
\over(|z+4ik_0t+\mu|^2+|\nu|^2)^2},
}
\tag1.5
$$
corresponding to the initial data
$$
q_0(z,0)=q_0(z)={2\overline\nu\over|z+\mu|^2+|\nu|^2}\exp\{k_0z
-\overline{k_0z}\},\tag1.6
$$
and the vanishing boundary conditions for the function
$g$ as $|z|\to\infty$.
\par
The discrete part of the scattering data is defined by frequencies,
amplitudes and phases of solitons. For the 1-soliton solution, it has the
following form:
$$
{\Cal L}^d=\{k_0,\mu,\nu\}\tag 1.7
$$
\par
If ${\Cal L}^d=\emptyset$, then the solution is called as nonsoliton.
\par
In the paper the asymptotics of the solution $q=q_\ve$ for the system (1.1)
with the initial data
$$
q_\ve(z,0)=q^\ve(z)=q_0(z)+\ve q_1(z),
\tag 1.8
$$
where $q_1$  is a smooth function with a finite support and
$\ve$  is  a small positive parameter, is constructed.
\medskip
{\bf 2. Solution asymptotics of scattering problem for $t=0$.}
Denote by $\phi_\ve$ the solution of (1.2), (1.3) for
$q=q_\ve$. The asymptotics of
$\psi^\ve(k,z)=\phi_\ve(k,z,0)$ is constructed in the form
$$
\psi^\ve(k,z)=\psi_0(k,z)+\ve\psi_1(k,z)+\dots. \tag 2.1
$$
\par
The problem (1.2), (1.3) is equivalent to the system
of the integral equations (\cite{7})
$$
(I-G[q,k])\phi=E_1,\tag 2.2
$$
where
$$
\eqalign{
&
G[q,k]\phi={i\over 4\pi}\int\int_{\Bbb C}d\zeta\wedge d\overline{\zeta}
\times\cr
&\times
\left(\matrix 0 &{q(\zeta,t)\exp\{k(z-\zeta)\}\over z-\zeta}\cr
-{\overline{q}(\zeta,t)\exp\{\overset{\hrulefill}\to{k(z-\zeta)}\}\over
        \overset{\hrulefill}\to{z-\zeta}} & 0
\endmatrix\right)\phi(k,\zeta,t),
}
$$
$I$ is the unit matrix, and
$E_1(kz)$ is the first column of the matrix  $E(kz)$.
\par
Substituting (2.1) and (1.8) into (2.2), we obtain the equations for
$\psi_0$ and $\psi_1$:
$$
\align
&
(I-G[q_0,k])\psi_0=E_1,\tag 2.3\\
&
(I-G[q_0,k])\psi_1=G[q_1,k]\psi_0,\tag 2.4
\endalign
$$
\par
The equation (2.3) has the explicit solution \cite{7}:
$$
\psi_0(k,z)=E_1(kz)-{\exp\{(k-k_0)z\}\over k-k_0}A_1(z),
\tag 2.5
$$
$$
A_{1}(z)={1\over \m z+\mu\m^2+\m\nu\m^2}
\left(
\matrix
\overset{\hrulefill}\to{(z+\mu)}\exp\{zk_0\}\cr
\nu \exp\{\overset{\hrulefill}\to{z k_0}\}
\endmatrix
\right).
$$
\noindent
The vectors $A_1$ and $A_2=\s\overline{A_1}$, where
$$
\s = \left(\matrix 0 & -1\cr
                                                1 & 0
                        \endmatrix\right),
$$
are the pair of the eigenfunctions for the equation:
$$
(I-G[q_0,k])W=F,\tag2.6
$$
as $k=k_0$.
\par
For the vectors $f$ and $w$, we define the sesquilinear form
$$
(f,w)_q=
\int\int_{\Bbb C} dz\wedge d\overline{z}\,
(\overline{q}\,f^{(1)}\overline{w^{(1)}}+
        q\,f^{(2)}\overline{w^{(2)}}),
\tag 2.7
$$
where $f^{(i)},\,w^{(i)}$ are the components of $f,\,w$.
\par
For the solvability of the equation (2.6)
as $k=k_0$ it is necessary to carry out the following condition:
$$
(F,C_i)_{q_0} = 0,\qquad i=1,2,\tag2.8
$$
where
$$
C_1(z)=
{1\over \m z+\mu\m^2+\m\nu\m^2}
\left(\matrix
\overset{\hrulefill}\to{(z+\mu)}\exp\{-z k_0\}\cr
\overline\nu \exp\{\overset{\hrulefill}\to{-z k_0}\}
\endmatrix
\right),\qquad
C_2=
\s \overline{C_1}
$$
are the eigenfunctions of the equation, which is formal adjoint for
(2.6) with respect to the sesquilinear form (2.7).
Hence, for the existence of the pole more than
the first order for the solution of  (2.4),
the sufficient condition is to carry out one of the following
inequalities:
$$
(G[q_1,k_0]\lim_{k\to k_0}\left((k-k_0)\psi_0\right),C_i)_{q_0}
\not= 0, \qquad i=1,2.
$$
Simple  transformations imply that the latter requirement is equivalent
to one of the inequality:
$$
Q_i=(A_1,C_i)_{q_1}\not=0,\qquad i=1,2.
$$
The constans $Q_i$ depend on the perturbation $q_1$.
We shall consider the function $q_1$ such that $Q_1\not=0$.
This perturbation we shall call the nondegenerate perturbation.
In this case the order of the pole for the correction
terms of (2.1) increases. It implies that this series is asymptotic as
$\vert k-k_0\vert>C \ve^\g$ (for any $0\le\g<1$), but become ineligible
at the neighborhood of $k=k_0$. By this reason, for
 $k$ close to $k_0$, we construct the asymptotics in the other form:
$$
\psi^\ve(k,z)=E(kz)\left(\ve^{-1}V_{-1}(\k,z)+
V_0(\k,z)+\dots\right), \tag 2.9
$$
where $\k=(k-k_0)\ve^{-1}$.
\par
Substituting (2.9) and (2.1) into (2.2), we obtain
the folowing equations:
$$
\align
(I-{\Cal G}[q_0,k_0])V_{-1}=
&
0,\tag 2.10\\
(I-{\Cal G}[q_0,k_0])V_0=
&
I+{\Cal G}[q_1,k_0]V_{-1}+
\k\pt_{k}{\Cal G}[q_0,k_0]V_{-1}+
\\
&+
\overline\k\pt_{\overline{k}}{\Cal G}[q_0,k_0]V_{-1},
\tag 2.11
\endalign
$$
where
$$
{\Cal G}[q,k]=
{i\over 4\pi}\int\int_{\Bbb C}d\zeta\wedge d\overline{\zeta}
\left(\matrix 0 &{q(\zeta,t)\exp\{\overline{k\zeta}-
k\zeta\}
\over z-\zeta}\cr
        -{\overline{q}(\zeta,t)\exp\{k\zeta -\overset{\hrulefill}
\to{k\zeta}\}\over
        \overset{\hrulefill}\to{z-\zeta}} & 0
\endmatrix\right).
$$
\par
One can see, that the function
$$
V_{-1}(\k,z)=E(-k_0z)A(z)\a(\k), \tag 2.12
$$
where $A$ is the matrix with the column $A_1$ and $A_2$,
is the solution of the homogeneous equation (2.10) for any vector $\a(\k)$.
\par
On the other hand, the solvability condition  (2.8) for (2.6)
implies that the solvability condition for the equation
$$
(I-{\Cal G}[q_0,k_0]){\Cal W}={\Cal F}
$$
has the form
$$
(E_1(k_0z){\Cal F},C_i)_{q_0}=0,\qquad i=1,2.
$$
\par
The vector $\a$ is determined from the solvability condition of
(2.11). Simple transformations and calculations give the following
quantities for the components of  $\a$:
$$
\eqalign{
\a^{(1)}(\k)
&=
-{4\pi i(\overline{Q_2+4\pi i\k})\over
        \m Q_1\m^2+\m Q_2+4\pi i\k\m^2},\cr
\a^{(2)}(\k)
&=
{4\pi iQ_1\over \m Q_1\m^2+\m Q_2+4\pi i\k\m^2}.}
\tag 2.13
$$
\par
Thus, for the nondegenerate perturbation ($Q_1\not=0$),
the asymptotic solution of  (1.2), (1.3)
has the form  (2.1), (2.5) as
$\vert k-k_0\vert>C \ve^\g$ and the form (2.9), (2.12), (2.13) as
$(k-k_0)=O(\ve^\delta)$ for any
$C>0$, $0\le\g<1$, $0<\delta\le1$.
Therefore,  for the nondegenerate perturbation of soliton
initial data, the combine solution of the scattering problem has no a pole
at the neighborhood of $k_0$.
\medskip
{\bf 3. Asymptotics of scattering data.}
The time evolution for the continuous component of the scattering data
is defined by the equality (\cite{7}):
$$
b(k,t)=b(k,0)\exp\{2it(k^2+\overline k^2)\}.
$$
So, for the calculation of its asymptotics, it is sufficient to
construct the asymptotics of  $b^\ve(k)=b(k,0)$.
Due to (2.5), (2.9) and (1.4) the asymptotics of $b^\ve$
has the composite form:
$$
\align
&
b^\ve(k)=\ve b_1(k)\,+\,\dots,\quad
\vert k-k_0\vert>C\ve^\g, \quad 0\le\g<1,\,C>0,\tag3.1\\
&
b^\ve(k)=
\ve^{-1}B_{-1}\left({k-k_0\over\ve}\right)
+B_0\left({k-k_0\over\ve}\right)+\dots,\quad
k-k_0=O(\ve^\delta),\quad 0<\delta\le1\tag 3.2
\endalign
$$
Moreover, the formula (1.4) and the asymptotics of $\psi^\ve$
imply the equality
$$
\align
\overline {b_1 (k)}
=&
-{i\over4\pi}\int\int_{\Bbb C} dz\wedge d\overline{z}\,\big(
\overline{q_0(z)}\,\psi_1^{(1)}(k,z)
\exp\{-\overset{\hrulefill}\to{kz}\}+\tag3.3
\\
&+
\overline{q_1(z)}\,\psi_0^{(1)}(k,z)
\exp\{-\overset{\hrulefill}\to{kz}\}\big)
\endalign
$$
The latter formula is not constructible becouse, for
$\psi_1$,  the clear expression is not obtained.
For  $b_1$, we can obtain the more effective formula.
\par
Direct calculation implies that the vector
$$
\psi^*_0(k,z)=
E_1(-kz)
-{\exp\{(-k+k_0)z\}\over k-k_0}
C_1(z)
$$
is the solution of the equation
$$
(I-G^*)\,\psi_0^*=E_1(-kz),
$$
where $G^*$ is the operator formally adjoint for
 $G$ with respect to the sesquilinear form (2.7).
\par
Computing value of the sesquilinear form $(\psi_1,\psi_0^*)_{q_0}$,
due to the equation (2.4) after simple transformations we
obtain that
$$
(\psi_1,E_1(-kz))_{q_0}+(\psi_0,E_1(-kz))_{q_1}=
(\psi_0,\psi_0^*)_{q_1},\tag3.4
$$
\par
The equalities (3.3) and (3.4) imply that
$$
b_1={i\over 4\pi}(\overline{\psi_0,\psi_0^*})_{q_1}
\tag 3.5
$$
\par
The equality (3.5) is similar to the formula for the variation of
the scattering data for the nonsoliton solution of the DS-II
equation from \cite{26}.
\par
For  $k-k_0=O(\ve^\delta)$, substituting (2.9) × (1.4), we
obtain that
$$
B_{-1}(\k)
\sim
-{i\over 4\pi}\int\int_{\Bbb C}
dz\wedge d\overline{z}\, q_0\,
\overset{\hrulefill}\to{V_{-1}^{(1)}}\exp\{\overline{kz}-kz\}
\tag 3.6
$$
\par
Further, substituting (1.6) and (2.12) in (3.6),
we come to the equality
$$
B_{-1}(\k)=-{\overline\a^{(2)}(\k)}.
$$
\medskip
\par{\bf 4. Asymptotics of solution for inverse problem in nonsoliton
case.} In the nonsoliton case the inverse scattering problem
(usually called the  $\bar D$-problem \cite{7}),
for the vector $u$ with the components
$$
u^{(1)}=\phi^{(1)}\exp\{-kz\}, \qquad
u^{(2)}=-\overline{\phi^{(2)}\exp\{-kz\}},
\tag4.1
$$
has the form
$$
\align
\left(\matrix \pt_{\bar k}& 0\cr
                                0 & \pt_k
                        \endmatrix\right)u
&=
\left(\matrix 0& -\overline {b(k,0)}\exp\{-is\}\\
        b(k,0)\exp\{is\}& 0\endmatrix\right)u,
\tag4.2\\
u
&\to \bigg(\matrix 1\cr
                                0\endmatrix\bigg),
\qquad |k|\to\infty,
\endalign
$$
where $s=2t(k^2+\bar k^2)-i(kz-\overline{kz})$.
\par
For $b(k,0)=b^{\ve}(k)$, the solution asymptotics of the problem (4.2)
is constructed in the form:
$$
u_\ve=u_0+\ve u_1+\dots.
\eqno(4.3)
$$
\par
The asymptotics of $b^{\ve}$ was constructed in the preceding section.
It has the composite form. The tipical variable for the principle
term of the asym\-p\-to\-tics reads as
$\k=(k-k_0)\ve^{-1}$. Therefore, it is naturally to come in
 (4.2) the variable of the same scale: $l=4\pi (k-k_0)\ve^{-1}$.
\par
Substitute (4.3), (3.2) in  (4.2) and pass to the variable $l$.
As a result, we obtain the
problems for $u_{j}$. The problem for the main term of the asymptotics
of the expansion (4.3) has the form:
$$
\align
\left(\matrix \pt_{\bar l}& 0\cr
                                0 & \pt_l
                        \endmatrix\right)u_0
&={1\over4\pi}
\left(\matrix 0& -\overline {B_{-1}}\exp\{-is\}\\
        B_{-1}\exp\{is\}& 0\endmatrix\right)u_0,\\
u_0
&\to \bigg(\matrix 1\cr
                                0\endmatrix\bigg),
\qquad |l|\to\infty.
\endalign
$$
\par
Due to passing to the variable $l$, the phase $s$ of the exponential contains
both the terms depending on the "fast" parameters  $z,\,t$,
and the "slow"
parameters $\ve t,\,\ve z$. It is naturally, the dependence on the "fast"
parameters to separate from the "slow" parameters. So, denote:
$$
S_0=2t(k_0^2+\bar k_0^2)-i(k_0 z-\overline{k_0 z}),
$$
$$
\theta={\ve\over4\pi}(4ik_0t+z),\quad \t={\ve t\over8\pi}.
$$
\par
After these exchanging, the problem for $u_0$ take the form:
$$
\eqalign{
\left(\matrix \pt_{\bar l}& 0\cr
                                0 & \pt_l
                        \endmatrix\right)u_0
&=
\left(\matrix 0& \overline {W_0}\exp\{-i\ve\t(l^2+\bar l^2)\}\cr
        -W_0\exp\{i\ve\t(l^2+\bar l^2)\}& 0\endmatrix\right)u_0,
\cr
u_0
&
\to \bigg(\matrix 1\cr
                                0\endmatrix\bigg),
\qquad |l|\to\infty.
}
\eqno(4.4)
$$
Here
$$
W_0={\overline{\O_1}\exp\{l\theta-\overline{l\theta}\}
\over |\O_1|^2+|\O_2+l|^2},
$$
$$
\O_1=iQ_1\exp\{-iS_0\},\qquad \O_2=iQ_2.
$$
\par
In system (4.4) the index of exponential, depending on
 $\t$, will be considered as the perturbation.
Extract two first terms in the Taylor expansion of this exponential
and the summand of the order $\ve$
carry over the equation for the next correction.
As a result of the such approximation, we obtain the problem for $u_0$:
$$
\align
\left(\matrix \pt_{\bar l}& 0\cr
                                0 & \pt_l
                        \endmatrix\right)u_0
&=
\left(\matrix 0& \overline {W_0}\\
        -W_0& 0\endmatrix\right)u_0,\\
u_0
&\to \bigg(\matrix 1\cr
                                0\endmatrix\bigg),
\qquad |l|\to\infty.
\endalign
$$
\par
For $\theta\not= 0$, the solution of this problem has the form:
$$
u_0(l,\theta,\O_1,\O_2)=
\left(\matrix 1\\
                0\endmatrix\right)-{1\over\theta}
                \left(\matrix \overline{(l+\O_2)}\\
                \overline{\O_1}\exp\{l\theta-\overline{l\theta}\}
                \endmatrix\right){1\over |\O_1|^2+|\O_2+l|^2}.
$$
\par
For $\theta\not=0$, the solution of the problem for the correction
$u_1$ exists
and is unique. However, $u_1$ has the pole of the second order with
respect to $\theta$ as $\theta=0$.
The expansion (4.3) is ineligible at the neighborhood of $\theta=0$
by the same reasons as the expansion (2.1) is ineligible
at the neigborhood of $k_0$.
\par
At a small neighborhood of $\theta=0$,
the asymptotic solution of the problem (4.2)
has the another form:
$$
u_\ve=\ve^{-1}U_{-1}+ U_0+\dots.
\eqno(4.5)
$$
\par
Substitute (4.5), (3.2) in (4.2). As a result, we obtain the problem
for $U_{-1}$:
$$
\align
\left(\matrix \pt_{\bar l}& 0\cr
                                0 & \pt_l
                        \endmatrix\right)U_{-1}
&=
\left(\matrix 0& \overline {\o_0}\\
        -\o_0& 0\endmatrix\right)U_{-1},\\
U_{-1}
&\to 0,
\qquad |l|\to\infty,
\tag4.6
\endalign
$$
where $\o_0={\overline{\O_1}\over |\O_1|^2+|\O_2+l|^2}$.
\par
The system (4.6) has two linearly independent
solutions decreasing as $|l|\to\infty$:
$$
{\Cal A}_1=\bigg(\matrix \overline{ \O_2+l}\cr
                        \overline{\O_1}\endmatrix\bigg)
{1\over\m l+\O_2\m^2+\m \O_1\m^2},\qquad
{\Cal A}_2=\bigg(\matrix  \O_1\cr
                        -(\O_2+l)\endmatrix\bigg)
{1\over\m l+\O_2\m^2+\m \O_1\m^2}.
$$
Hence, the general solution of (4.6) decreasing at infinity reads as
follows:
$$
U_{-1}={\Cal A}\b(\theta),
$$
where ${\Cal A}$ is the matrix with the column
${\Cal A}_j$ and  $\b(\theta)$ is an arbitrary vector.
\par
For calculation $\b$, we consider the problem for $U_{0}$.
Taking in consideration the term carrying over the previous step,
it has the form:
$$
\eqalign
{
&\left(\matrix \pt_{\bar l}& 0\cr
                                0 & \pt_l
                        \endmatrix\right)U_{0}
=
\left(\matrix 0& \overline {\o_0}\cr
        -\o_0& 0\endmatrix\right)U_{0}
+
\left(\matrix 0& \overline {\o_1}\cr
        -\o_1& 0\endmatrix\right)U_{-1}+
\cr
&+
\left(\matrix 0& \overline {\o_0}[(\overline{lZ}-lZ)-i\t(l^2+\bar l^2)]\cr
        -\o_0[(lZ-\overline{lZ})+i\t(l^2+\bar l^2)]& 0\endmatrix\right)
        U_{-1},
\cr
&
U_{0}\to \left(\matrix 1\cr
                0\endmatrix\right),
\quad |l|\to\infty,
}
\tag 4.7
$$
where $\o_1=\,-\,{B_0\over4\pi}\exp\{iS_0\}$, $Z={1\over4\pi}(4ik_0 t+z)$.
The system of the integral equations corresponding to (4.7)
reads as follows:
$$
(I-{\Cal H}[\o_0,l])U_0=\left(\matrix 1\cr
                        0\endmatrix\right)+
{\Cal H}[\o_1,l]U_{-1}+{\Cal H}[f,l]U_{-1},
\tag 4.8
$$
where
$$
{\Cal H}[h,l]w=\,-\,
{i\over2\pi}
\int\int_{\Bbb C}dm\wedge d\overline{m}
\left(\matrix 0 &{\overline{h}(m)\over m-l}\cr
-{h(m)\over\overline{m-l}} & 0
\endmatrix\right)w(m),
$$
$$
f=[(lZ-\overline{lZ})+i\t(l^2+\bar l^2)]\o_0.
$$
\par
The solvability condition for (4.8) has the form:
$$
({\Cal B}_1,F)_{\o_0}=0,\quad ({\Cal B}_2,F)_{\o_0}=0,
\tag4.9
$$
where $F$ is the right hand side in (4.8),
$$
{\Cal B}_1=\bigg(\matrix \overline{ \O_2+l}\cr
                        \O_1\endmatrix\bigg)
{1\over\m l+\O_2\m^2+\m \O_1\m^2},
\quad
{\Cal B}_2=\s\overline{{\Cal B}_1}
$$
and the integration in the sesquilinear form is realized with respect to
 $l$.
Note that $({\Cal B}_1,{\Cal A}_2)_{\o_0}=
({\Cal B}_2,{\Cal A}_1)_{\o_0}=0$.
\par
The solvability condition (4.9) implies the system of the equations for
 $\b$:
$$
\eqalign{
& M_1\overline{\b^{(1)}}+ (\overline{M_2}+2i\pi
Z-\pi\t\O_2)\overline{\b^{(2)}}=0,
\cr
& (M_2-2i\pi\bar Z-\pi\t\overline{\O_2})\overline{\b^{(1)}}-
\overline{M_1}\overline{\b^{(2)}}=-2i\pi. }
\tag4.10
$$
Here
$$
M_1=({\Cal B}_1,{\Cal A}_1)_{\o_1},\qquad
M_2=({\Cal B}_2,{\Cal A}_1)_{\o_1},
$$
and the integration in the sesquilinear form is realized with respect to
 $l$. From (4.10), the formulae for the components of $\b$ are deduced:
$$
\eqalign{
\b^{(1)}
&=
{2i\pi(M_2-2i\pi Z-\pi\t \O_2)
\over
\m M_1\m^2+\m M_2+2\pi Z-\pi\t\bar \O_2\m^2},
\cr
\b^{(2)}
&= {-2i\pi\overline{M_1}
\over
\m M_1\m^2+\m M_2+2\pi Z-\pi\t\bar \O_2\m^2}.
}
$$
\par
Taking in consideration the dependence on $S_0$ for $\o_1$ and the
explicit form of the vectors ${\Cal B}_{1,2}$, we correct $M_{1,2}$:
$$
M_1=\l_1\exp\{-iS_0\},\quad {\hbox{where}}\quad \l_1={\hbox{const}},
\quad M_2={\hbox{const}}.
\tag4.11
$$
\par
Thus, the asymptotical solution of (4.2) has the form (4.3)
as $\theta>\ve^{\d}$,  $0\le\d<1$, and the form (4.5) as
$\theta\le\ve^{\g}$, $0<\g\le 1$.
\medskip
\par{\bf 5. Asymptotics of solution for DS-II equation.}
The solution of the DS-II equation can be constructed
by the following formula:
$$
q(z,t)
={i\over \pi}\int\int_{\Bbb C}dp\wedge d\overline{p}
b(p,t)\,u^{(1)}(z,t,p)\exp\{is\}.
\tag5.1
$$
Using the constructed asymptotics of the scattering data (3.1), (3.2) and
the asym\-p\-to\-ti\-cal representation (4.3), (4.7) of
$u^{(1)}$, we shall construct   the principle term of the
formal asymptotics for $q_\ve$. The structure of the asymptotical expansion
for  $u^{(1)}$ implies, that the asymptotics of
 $q_\ve$ is different for different $\theta$.
\par
If $\theta>O(\ve^{\d})$ ($\d<1$), than, for the calculation of $q_\ve$
we use the formulae (3.1), (3.2) and (4.3).
The analize of the asymptotics for
the integral (5.1) with respect to $\ve$ implies that
$$
q_\ve=O(\ve^{1-\d}), \quad{\hbox{as}}\quad\theta>O(\ve^{\d}),\quad\d<1.
\tag5.2
$$
\par
If $\theta\le\ve^{\d}$, than, for the calculation of $q_\ve$,
we substitute (3.1), (3.2), (4.7) in (5.1). As a result we obtain that
$$
q_{\ve}(z,t)
\sim
{i\over 16\pi^3}\int\int_{\Bbb C}dl\wedge d\overline{l}
B_{-1}(l)\,U^{(1)}_{-1}(l,z,t)\exp\{iS_0\}\,+\,O(\ve).
$$
The latter formula implies:
$$
q_\ve(z,t)
\sim
{-i\overline{M_1}
\over
\m M_1\m^2+\m M_2+{1\over2}(4k_0 t+z)-\pi\t\bar \O_2\m^2}.
\tag5.3
$$
\par
The constants $\l_1$ and $M_2$ (see (4.11)) can be determined putting
$t=0$ in (5.3). Due to (5.3) and (1.6) we obtain that
$$
\l_1=-i{\nu\over2},\quad M_2={\mu\over2}.
$$
\par
The comparising of the parameters for the perturbed solution (5.3) and the
pure soliton solution (1.7) shows that the frequency $k_0$ of the soliton is
invarible under the perturbation of the initial data and its phase shift is
modulated by time in the form:
$$
\mu_{\ve}(\t)=\mu-2\pi\t\O_2,
$$
Thus, for $0<t<\ve^{-1}\,{\hbox{Const}}$, the asymptotical solution of
(1.1), (1.8) reads as follows:
$$
\eqalign{
q_{\ve}(z,t)
&
\sim
{2\overline{\nu}
\over|z+4ik_0t+\mu_{\ve}|^2+|\nu|^2}
\exp\{k_0z-\overline{k_0z}+2i(k_0^2+\overline k_0^2)t\},
\cr
g_{\ve}(z,t)
&\sim
{-4\overline{(z+4ik_0 t+\mu_{\ve})}^2
\over(|z+4ik_0t+\mu_{\ve}|^2+|\nu|^2)^2}.
}
$$

\enddocument
\end